\documentclass[%
 reprint,
 amsmath,amssymb,
 aps,
]{revtex4-1}

\usepackage{graphicx}
\usepackage{dcolumn}
\usepackage{bm}

\begin{document}

\preprint{APS/123-QED}

\title{Study of the phase diagram of evaporation-condensation systems with a Histogram Reweighting adaptation method}

\author{G. J. dos Santos}
 \altaffiliation[Also at ]{Departamento de F\'{\i}sica, Instituto de F\'{\i}sica Aplicada, Universidad Nacional de San Luis-CONICET, Chacabuco 917, D5700HHW, San  Luis, Argentina}
\author{D. H. Linares}%
\affiliation{Departamento de F\'{\i}sica, Instituto de F\'{\i}sica Aplicada, Universidad Nacional de San Luis-CONICET, Chacabuco 917, D5700HHW, San  Luis, Argentina}%

\author{A. J. Ramirez-Pastor}%
\affiliation{Departamento de F\'{\i}sica, Instituto de F\'{\i}sica Aplicada, Universidad Nacional de San Luis-CONICET, Chacabuco 917, D5700HHW, San  Luis, Argentina}%


\date{\today}

\begin{abstract}
The critical point of the condensation transition for linear molecules adsorbed on square lattices, was studied by using an adaptation of the Histogram Reweighting technique. The results were obtained by means of grand canonical Monte Carlo simulations within the lattice gas model, along with finite size scaling using the fourth order Binder cumulant. The Method was tested in a system of interacting monomers in which the critical point can be determined exactly. The application of this method to the determination of the critical point in dimer systems with attractive interactions, gave better results than the previous reported studies to the best knowledge of the authors. In addition, the adsorption isotherms at different temperatures, as well as the phase diagrams for monomer and dimer systems were obtained, achieving significant improvements in the phase diagram for dimers. 
\begin{description}
\item[PACS numbers]
68.43.-h, 
68.35.Rh,   
64.60.De,    
47.11.Qr, 
87.55.kh, 
74.62.-c 
\end{description}
\end{abstract}

\pacs{Valid PACS appear here}
\maketitle


\section{\label{sec:level1}Introduction}

he statistics and thermodynamics of adsorbed dimers have been one of the most interesting problems in surface science and statistical mechanics \cite{Fowler,Kaste1,Kaste2,Fisher,Nagle,Lieb,Phares,SS3,PRB4,JCP16,PRB8,RB33,RB34,RB35,RB36,RB37,RB38,RB39,RB40}. An early seminal contribution to dimer statistics was made by Fowler and Rushbrooke \cite{Fowler}, while an exact solution was found by Kasteleyn \cite{Kaste1,Kaste2}, but only at close-packed density.


At intermediate coverage, an intriguing problem is the ordering of repulsive dimers on various lattices. Phares et al. \cite{Phares} studied interacting dimers on a semi-infinite square lattice using the transfer-matrix method. The authors concluded that there was a finite number of ordered structures for dimers with repulsive nearest-neighbor interactions. Later, the simulation analysis of the phase diagram of dimers with repulsive nearest-neighbor interactions on a square lattice \cite{SS3} confirmed the presence of the two well-defined structures: a $(4 \times 2)$ ordered phase at $\theta =1/2$ and a ``zigzag" (ZZ) order at $\theta =1/2$, being $\theta$ the surface coverage.

In a later work \cite{PRB4}, from the $(4 \times 2)$ phase appearing in dimers at half coverage, it was possible (1) to predict the existence of a $(2k \times 2)$ structure for $k$-mers at half coverage and (2) to obtain the critical temperature $T_c(k)$ characterizing the transition from the disordered state to the $(2k \times 2)$ phase as a function of the size $k$ of the adsorbed molecules. Ref. \cite{JCP16} went a step further, analyzing the universality class of the phase transition at $\theta =1/2$. By using Monte Carlo (MC) simulations and finite-size scaling (FSS) analysis, the critical behavior of the system was studied. The obtained results indicated that the nature of the phase transition occurring at half coverage in a system of repulsive rigid $k$-mers on a square lattice changes from second order for $k = 1$ to first order for $k \geq 2$.

The ZZ order corresponding to repulsive dimers on square lattices at 2/3 coverage was studied in Ref. \cite{PRB8}. The calculations were performed by using exchange MC simulations and FSS theory. The exhaustive determination of the complete set of static critical exponents, along with the behavior of Binder cumulants, confirmed previous results in the literature \cite{Phares,SS3} namely, the existence of a continuous phase transition at 2/3 coverage. Although it was not possible to exclude the existence of a more complex critical behavior, the results suggest that the phase transition does not belong to the universality class of the two-dimensional Ising model.


In the case of attractive lateral interactions, the phase diagram corresponding to homonuclear dimers adsorbed on square lattices was studied by using MC simulations \cite{SS3}. For temperatures below the critical value, the system undergoes a first-order phase transition which was observed in the clear discontinuity in the adsorption isotherms. The critical temperature was obtained from the measurement of the maximum in the specific heat, which was calculated in the canonical ensemble for each surface coverage $\theta$ in the range $[0,1]$. The resulting phase diagram is similar to that reported for a simple lattice gas of monomers (or equivalently, an Ising ferromagnet) with the critical temperature shifted to a higher value by a factor of $\approx 1.2$. The critical temperature for the simple lattice gas is given by $k_BT_c /w \approx 0.56725$ \cite{Hill} ($k_B$ is the Boltzmann constant, $T_c$ is the critical temperature and $w$ is the magnitude of the lateral interaction energy).

By using MC simulations, multiple-histogram reweighting and finite size-scaling techniques, R\.zysko and M. Bor\'owko \cite{RB33,RB34,RB35,RB36,RB37,RB38,RB39,RB40} studied a wide variety of systems in presence of multisite occupancy. Among them, attracting dimers in the presence of energetic heterogeneity \cite{RB33}, heteronuclear dimers consisting of different segments, A and B, adsorbed on square lattices \cite{RB34,RB35,RB36,RB37,RB38}, and trimers with different structures adsorbed on square lattices and pores \cite{RB39,RB40}. In these papers, a rich variety of phase transitions was reported along with a detailed discussion about critical exponents and universality class. In the particular case of attractive homonuclear dimers on square lattices, the authors corroborated the results previously obtained in Ref. \cite{SS3}.


The attractive phase diagram reported in Ref. \cite{SS3} was obtained from the extrapolation of $T_c(L)$ towards the thermodynamic limit, where $T_c(L)$ is the temperature of the maximum in the specific heat for a $L \times L$ lattice. This quantity scales with the size of the system as $L^2$ \cite{Binder}. The calculations were performed in the canonical ensemble, using the coverage as the control parameter. This procedure does not allow to simultaneously obtain critical coverage and critical temperature. In fact, the value of $T_c$ reported in Ref. \cite{SS3} was calculated at 1/2 coverage, and not at the critical coverage (which is expected to be shifted from 0.5). In addition, recent results in the literature \cite{Almarza,Lopez} show that fixing the density in models such as that discussed here, corresponds to introducing a constraint that renormalizes the critical parameters characterizing the phase transition.

To remedy this situation, we perform new simulations in the grand canonical ensemble, using the chemical potential as control parameter. The primary objective of this paper is to provide a detailed description of this new scenario, revisiting the classical problem of attractive dimers on square lattices. In order to do so, extensive MC simulations in the grand canonical ensemble, based on an efficient histogram reweighting (HR) technique and using high-performance computational capabilities, have been carried out.

 All the quantities measured in Ref. \cite{SS3} were recalculated. In addition, the study was complemented with new measurements such as density distribution, energy distribution and order parameter cumulant over a wide range of temperatures. The obtained results indicated that the critical coverage is shifted to a lower value than 0.5, resulting in a slightly asymmetric phase diagram. In the same way, the critical temperature is found to be  $T_c=0.692$ which is a higher value than the one mentioned previously. This technique was tested succesfully contrasting  the results with the exact  solution of the monomer adsorption system and the results obtained in the attractive dimers problem, clearly corrects and complete the ones obtained in Ref. \cite{SS3}.

This paper is organized as follows. The model is described in Section 2. Details of the Monte Carlo simulations and HR technique are presented in section 3. The developed methodology which is based in an adaptation of the HR technique applied to the fourth order Binder cumulant is detailed in Section 4. Finally, in Section 5 the results and conclusions are presented.

\section{\label{sec:level1}Model}

The system consisting of linear molecules adsorbing on a flat surface (see Fig. 1), was modelled by $k$-mers adsorbing on a two-dimensional square lattice of linear size L with periodic boundary conditions. A $k$-mer of size $k$, consists of $k$ identical consecutive segments, each one occupying exactly one lattice site. The distance between $k$-mer segments is equal to the lattice constant, so each adsorbed $k$-mer will occupy $k$ consecutive lattice sites and may be adsorbed in two directions ($x$ and $y$). Each segment of a $k$-mer interacts only with nearest neighbouring segments of other $k$-mers with interaction energy $w$. In order to characterize the occupancy state of each lattice site the occupancy variable $c_i$ was introduced, so that when site $i$ is empty (occupied) $c_i=0$ ($c_i=1$). Then, the generalized grand canonical Hamiltonian is,

\begin{equation*}
  H=\sum_{<i,j>}^{} w c_i c_j  - N(k-1)w + \varepsilon_0\sum_{i}^{} c_i
\end{equation*}

where $w$ is the nearest-neighbour (NN) interaction and it is considered attractive (negative) and $<i,j>$ denotes pairs of NN sites. This summation overestimates the number of pairs by $N(k-1)$, so the middle term $N(k-1)w$ corrects the total interaction energy. Finally, $\varepsilon_0$ represents the adsorption energy of one site, but since the surface is considered homogeneous this energy has been neglected. It is necessary to mention that $w$ is expressed in $k_B T$ units.

\begin{figure}
\hspace*{-0.65cm}
\includegraphics[scale=0.3]{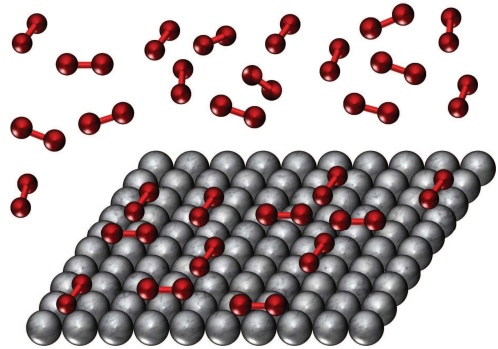}

\caption{Schematic representation of the adsorption-desorption system for the case of dimers $k=2$.}
\end{figure}

\section{\label{sec:level1}Monte Carlo and Histogram Reweighting Simulations}

The problem was studied by means of grand canonical Monte Carlo simulations using a Parallel-Tempering algorithm \cite{HPT-1-Swendsen,earl2005parallel,HPT-3-hukushima}, that allows the system to reach equilibrium in a considerable shorter time than the standard Monte Carlo simulations. The Parallel Tempering is applied on chemical potential ($\mu$), generating a collection of $N_{rep}$ replicas, each one at a different value of $\mu$ ranging from $\mu_0$ to $\mu_f$ where, $N_{rep}$, $\mu_0$ and $\mu_f$ are parameters of the simulation. The algorithm selects one of the $N_{rep}$ replicas at random and then a set of $k$ consecutive sites (linear $k$-uple) is chosen randomly in that particular replica. Then, the operating dynamics consists in an attempt to change the occupancy state of the selected $k$-uple: if it is an "empty $k$-uple", adsorption is followed with probability $P_{ads}=min\{1,exp(-\Delta H/k_B T+\mu \Delta N/k_B T)\}$, and if it is an occupied $k$-uple,  desorption is tried out with probability $P_{des}=min\{1,exp(-\Delta H/k_B T-\mu \Delta N/k_B T)\}$. Where $\Delta H$ is the difference between the Hamiltonians  and $\Delta N$ is the difference in the number of adsorbed $k$-mers respectively  between the initial and final states. It is worth mentioning that the chemical potential $\mu$ is expressed in $k_B T$ units, where $k_B$ is the Boltzmann's constant. 

After $M=L \times L$ attempts of changing different $k$-uple states, an attempt of interchanging configurations between neighbouring replicas is tried out with probability $P=min\{1,exp(-\Delta N\Delta \mu)\}$, where $\Delta N$ is the difference in the number of molecules and $\Delta\mu$ is the difference in chemical potential between the two interchanging replicas. 

A Monte Carlo Step (MCS) consist of $L \times L$ attempts per replica of changing a $k$-uple occupation state, i.e. $MCS=L \times L \times N_{rep}$, where the linear dimensions of the system $L$ ranges between 20 and 120.

For each replica a random initial configuration is generated, and it was checked that equilibrium is reached after $r=2 \times 10^6$ MCS; the next $2 \times 10^6$ MCS where used to compute averages. 

The Parallel Tempering algorithm is used mainly to unblock freezing states using a given number of replicas of the system. The number of replicas in the simulations, must be one such that the exchange probability is large, typically greater than 0.5. On the other hand, this number should not be too large to compromise the calculation time. Given the above, the simulations of the present work were run with a hundred replicas ($N_{rep}=100$).

Typical quantities monitored in the simulations are the surface coverage $\theta$ and energy per site $\epsilon$ which are calculated as simple averages,

\begin{equation}\label{}
  \theta= \frac{1}{M} \sum_{i}^{M} <c_i>
\end{equation}

\begin{equation}\label{}
  E=\frac{1}{M} <H>
\end{equation}

where $<...>$ means time average over the Monte Carlo simulation.

Histogram reweighting methods are great tools for extracting as much information as possible from a single simulation. This technique is well detailed elsewhere \cite{panagiotopoulos2000monte,ferrenberg1988new,ferrenberg1989optimized} , so here we will give a short explanation of the subject.
Near first or second order phase transitions, some thermodynamical quantities or their derivatives, such as specific heat or, in our case, the lattice coverage, show pronounced peaks or discontinuities. This effect is a big obstacle when trying to obtain results from direct simulations. In condensation transitions, such as in this work, the isotherms shows a sharp jump between the two states (Fig. 2) near the critical temperature.

\begin{figure}

\includegraphics[scale=0.35]{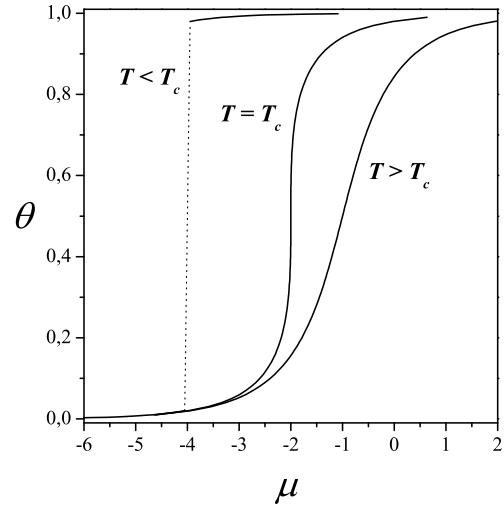}
\caption{Schematic representation of the typical behaviour of the adsorption isotherms for different temperatures near a first-order phase transition. In finite systems, the discontinuity shown for $T<T_c$ is actually a sharp jump.}
\end{figure}

This jump makes it very difficult to obtain information of the system in that region from a direct simulation, since little variations in the chemical potential lead to very large changes in the lattice coverage. In addition, due to the fluctuations, it is necessary to take a lot of samples in order to minimize the statistical errors, which has a high computational cost. Here is when the Single-HR technique is very useful.

From a grand canonical simulation of $J$ steps run at $\mu=\mu_0$ and $T=T_0$, one can create a two-dimensional histogram $H_{T_0,\mu_0}(U,N)$, where $U$ is the total energy and $N$ is the number of molecules. The probability distribution is related to the histogram in the following way,

$P(U,N,T_0,\mu_0)=H_{T_0,\mu_0}(U,N)/J$

Single-HR technique allows us to obtain the probability distribution at a slightly different $\mu$ and $T$ from,

\begin{equation}\label{2}
\resizebox{.9\hsize}{!}{$P(U,N,T,\mu)= \frac{H_{T_0,\mu_0}(N,U)exp[-(\beta-\beta_0)U+N(\beta\mu-\beta_0\mu_0)]}{\sum_{N}\sum_{U}H_{T_0,\mu_0}(N,U)exp[-(\beta-\beta_0)U+N(\beta\mu-\beta_0\mu_0)]}$}
\end{equation},


where $\beta=1/(k_B T)$.

This is a very powerful relationship, since the average values of any function of N and U can be estimated from,

\begin{equation}\label{promedios}
<A>=\sum_{N}\sum_{U} A(U) P(N,U,T,\mu)
\end{equation}

\section{\label{sec:level1}Methodology: Histogram Reweighting for Binder Cumulant Curves}

In order to determine the condensation critical temperature, the fourth order Binder cumulant \cite{muller1979monte} of the order parameter $m$ was calculated using HR techniques, where the Binder cumulant is defined as,  

\begin{equation}\label{eq1}
<U>_L = 1- \frac{<m>^4}{3<m^2>^2},
  \end{equation}

where $<...>$ denotes average over configurations, and $m=\theta-<\theta>$ is the standard order parameter for liquid-vapor transition. Plotting $U_L$ vs $T$ for different system sizes ($L$), the curves will meet in an intersection point where the cumulant takes a fixed universal value $U^*$. The temperature at which this intersection occurs, is the critical temperature. Therefore, we are looking for plots of $U_{L}$ vs $T$, as the one shown in Fig. 3.

The simulations were performed in the grand canonical ensemble at different temperatures, so $m_L$ as well as $U_L$, are functions of $T$ and also functions of $\mu$. The plot in Fig. 3 is a projection of a ($\mu,T,U_L$) graphic onto a ($T,U_L$) plane. 

\begin{figure}[]

\includegraphics[scale=0.35]{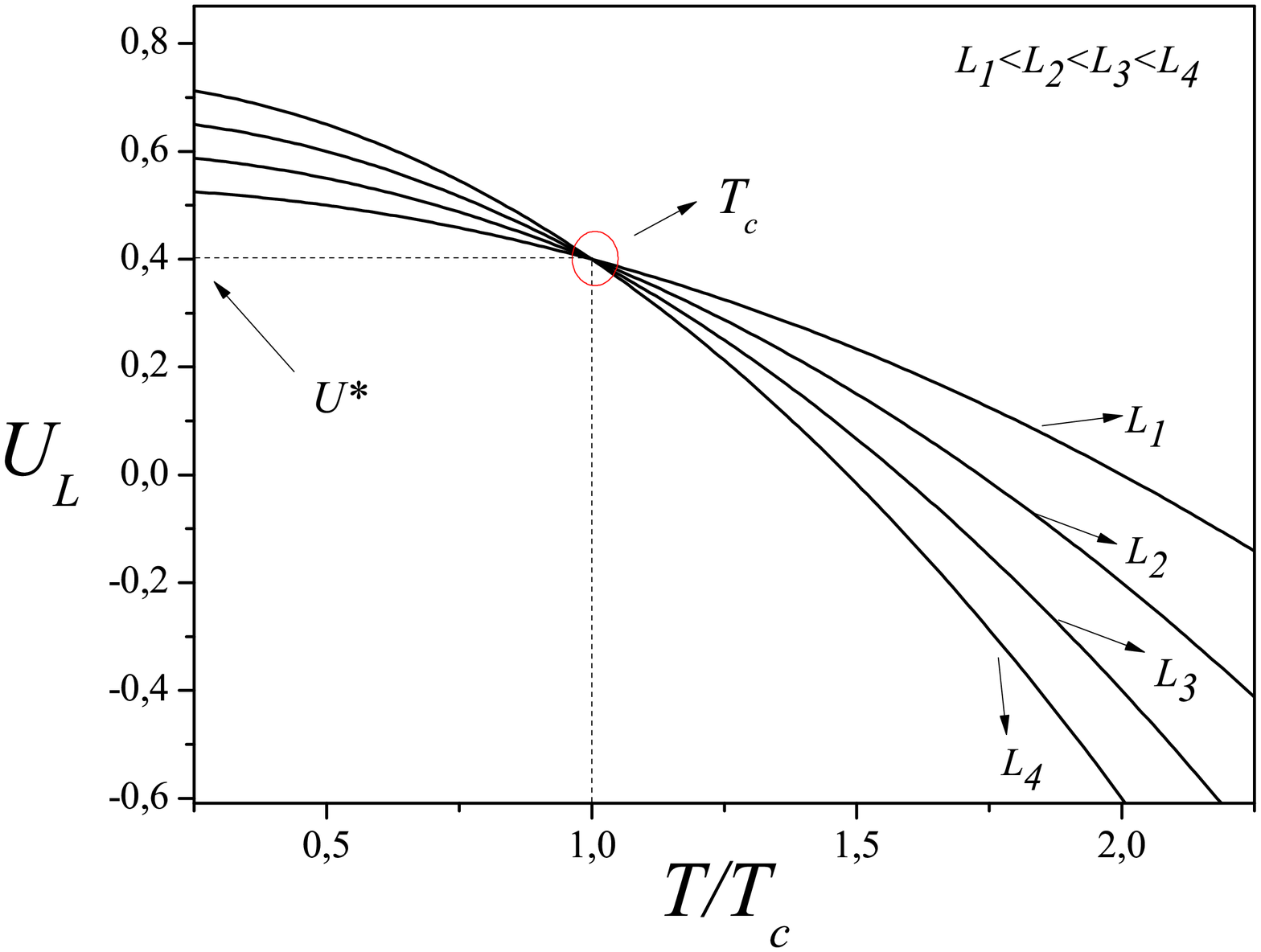}
\caption{Schematic representation of a typical Binder cumulant intersection plot. The curves correspond to the fourth order Binder cumulant plotted against temperature for different system sizes L. The intersection point of the curves determines the critical temperature and the fixed universal value of the cumulant $U*$ .}
\end{figure}

The question that naturally arises is, which are the values of $\mu$  that give the correct intersection plot projection. At low temperatures the isotherms show a discontinuity for a given $\mu$ that highlights the condensation phase transition occurring in the system (Fig. 2). As the temperature increases, the size of the discontinuity decreases. At the precise temperature at which this discontinuity becomes a point, the system goes from experiencing a first order phase transition to experiencing a second order phase transition; being the system at the so-called critical point. Correspondingly, the isotherms go from having a discontinuity to having an inflection point. It is for this reason that if we move in the plane ($\mu$;$T$) following a trajectory $Z$ ($\mu$,$T$) = 0 (or $\mu = h(T)$) defined by the inflection point of the different isotherms, we will eventually encounter the critical point as shown in Fig. 4. We can evaluate the $U_L$ cumulant along this trajectory finding a three-dimensional curve in a space ($\mu$,$T$, $U_L$). Due to the scale invariance of $U_L(T_c)$, if we repeat this procedure for the different sizes of the system $L$, we will find a family of curves that intersect at a given point, "the critical point". Due to the previously established relationship between $\mu$ and $T$, $Z(\mu,T)=0$, we can detach from a degree of freedom and work with a single independent variable: $T$. Therefore, to find $T_c$, it would be enough to draw the curves in the bidimensional space ($\mu(T)$,$T$, $U_L$) for each $L$, such as those shown in Fig. 3, and find the intersection point.

\begin{figure}

\includegraphics[scale=0.35]{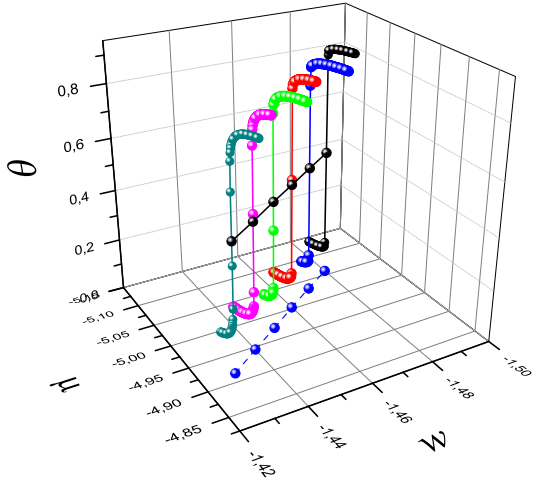}
\caption{3D plot of a set of isotherms at different temperatures. It can be seen the trajectory followed in the $(\mu, w)$ plane in order to obtain the critical temperature, corresponding to the projection of the inflexion points. From front to back the isotherms correspond to $w =-1.43, -1,44, -1.45, -1.46, -1.47, -1.48$ in $k_B T$ units.}
\end{figure}

In order to find the location of the inflection point for a given isotherm, the second derivative of the isotherm should be obtained. It is known that to obtain an accurate enough derivative of a function, it is necessary that the primitive consists of a very large number of points, so the case of a second derivative requires even more points, demanding a very large computational cost since the isotherms typically show very sharp increases near first-order phase transitions. On the other hand, it can be seen in Fig. 5 that over an isotherm (fixed $T$) the fourth order Binder cumulant experiences a maximum at the same point where the isotherm's inflection point occurs, giving us a method to obtain the value of $\mu^*(T)$ and hence the value of $U_L(\mu^*(T),T)$. Repeating this procedure for a group of isotherms of a system of size $L$, we can obtain a $U_L(T)$ curve. If the procedure is repeated for the different system sizes $L$, we will obtain a family of curves $U{_{L_i}}(T)$ whose intersection point determines the critical temperature $T_c$.

\begin{figure}

\includegraphics[scale=0.35]{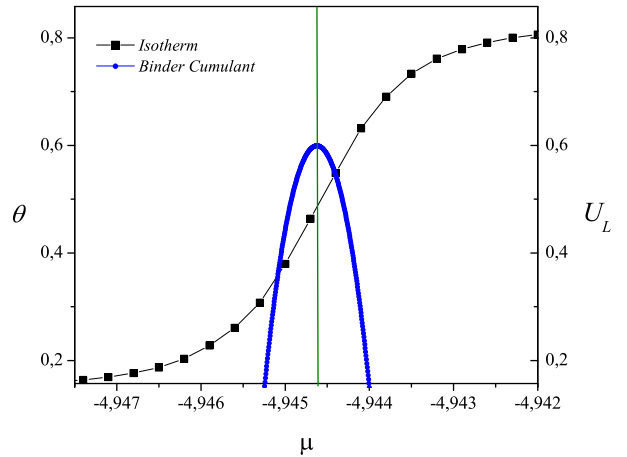}
\hspace*{-2cm}
\caption{Dimers adsorption isotherm on a square lattice, along with the fourth order Binder cumulant plotted against the chemical potential $\mu$ with fixed $w=-1,44$. The straight line indicates the location of the inflection point of the isotherm, showing that it occurs when the cumulant takes its maximum value.}
\end{figure}

This methodology is further detailed at next: \begin{itemize}
                            \item For a given temperature a histogram is obtained directly from the simulation at a chemical potential as near as we can find it from the inflection point.
                            \item Using HR technique the cumulant $<U>_L = 1- \frac{<m>^4}{3<m^2>^2}$ is calculated for different values of $\mu$ near the original one, tuning the chemical potential.
                            \item  The maximum value of $U_L$ is taken from a $U_L$ vs $\mu$ plot.
                            \item  We repeat this sequence for different temperatures and finally build the $U_L$ vs $T$ plot for different lattice sizes ($L$) and find the intersection point.
                          \end{itemize}

Once we have found the critical temperature, we use HR technique one more time to calculate the critical isotherm. Finding the inflection point of this isotherm we find the critical coverage ($\theta_c$).

In order to obtain the phase diagram, the two-state approximation was used \cite{HPT-4-2stateaprox} along with HR technique. For a given isotherm, we construct a histogram for a value of the chemical potential as near as possible to the inflection point. Now,  the reweighting method is applied to this histogram, then the chemical potential is tuned until the areas under the two peaks becomes equal. The two density values at which this peaks occur are the equilibrium points in the coexistence curve, see Fig. 6. 

\begin{figure}

\includegraphics[scale=0.35]{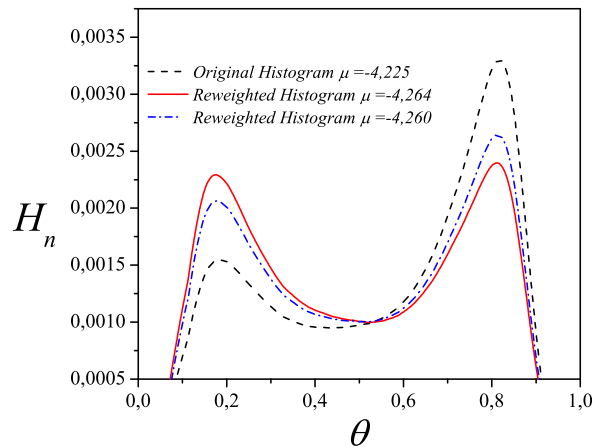}
\caption{Reweighted lattice coverage histograms for dimers adsorption at fixed $w=-1,39$. The original histogram is obtained directly from the simulation and corresponds to a chemical potential $\mu=-4.225$. Applying histogram reweighting technique, the chemical potential is "tunned" until the resulting histogram curve is such that the areas under the two peaks is equal. The coverage values of the two maximum are the corresponding coverage values in the phase diagram for that given $w$ or temperature.}
\end{figure}

\section{\label{sec:level1}Results and Conclusions}

Monte Carlo simulations along with the HR technique were performed to determine the critical temperature in two different "rod-like" molecule condensation-evaporation systems. In order to test the methodology described in the last section, the case of monomer ($k$=1) adsorption on square lattices was studied. It is well known that this problem has an exact solution \cite{Libro-Hill-2}, making it ideal for testing the accuracy of the method. For this purpose, we ran MC simulations for attracting monomers in different lattice sizes $L=50,60,80,100,120$ and with various interaction energies $w=-1.70, -1.71 ...-1.80$. The typical behavior of the adsorption isotherms at different values of $w$ is shown in Fig. 7.

\begin{figure}

\includegraphics[scale=0.35]{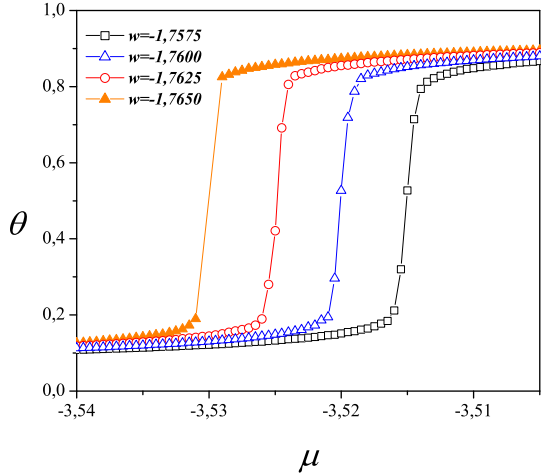}
\caption{Monomers adsorption isotherms on a square lattice of size $M=100 \times 100$ for diferent values of $w$. It can be seen the typical behaviour of the isotherm when the system experiences a first order phase transition for $w<-1,7625$. The diferent curves correspond to $w=-1,7575$ (empty squares), $w=-1,7600$ (empty triangles), $w=-1,7625$ (empty circles), $w=-1,7650$ (filled triangles).}
\end{figure}

At first, the accuracy of the method was tested contrasting the value of the critical temperature given by the exact solution and the value obtained from the methodology used here. From the two-dimensional histograms $H(N,U)$ obtained directly from the simulations employing the procedure described in section (4) the fourth order Binder cumulant was calculated  as a function of temperature $U_L(T)$ for different lattice sizes as shown in Fig. 8.

\begin{figure}

\includegraphics[scale=0.35]{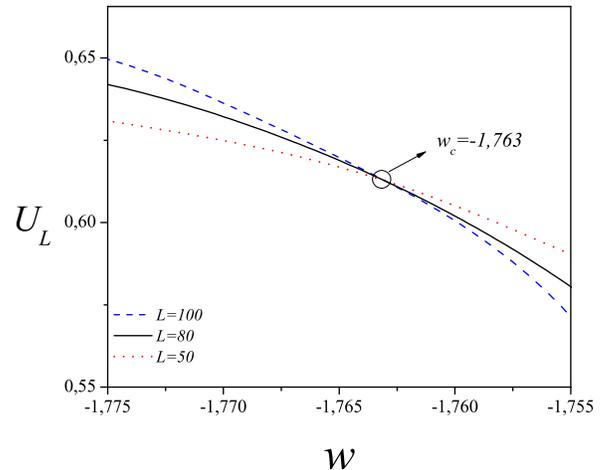}
\caption{Intersection plot of the fourth order Binder cumulant for the monomer adsorption system. The critical interaction energy found $w_c=-1,763$, is related to the critical temperature by $T_c=1/w_c=0.567$.}
\end{figure}

It can be seen from Fig. 8, that the cumulants $U_L(T)$ corresponding to different lattice sizes ($L$) intersect at a well defined point, giving a value of the critical temperature of $T_C=0,56721$.  Since the exact value is $T^*_c=0,56725$, it can be concluded that the methodology is accurate enough, to be used in this type of systems. In addition, the critical density of this system was calculated by the methodology described previously, finding the value $\theta_c=0,501$. Again, this result is in excellent agreement with the exact solution, and provides further confidence and evidence that the methodology presented here is valid to find characteristic behaviours of the critical point. 

Once the validity and the accuracy of the methodology were tested, the study of a system of attracting homonuclear dimers ($k=2$) adsorbed on square lattices was carried out. The objective in this case, is to revisit the study realized by Ramirez-Pastor et.al \cite{SS3} where the critical temperature of the phase transition of such a system was determined along with the corresponding phase diagram. In this work, the authors used canonical Monte Carlo simulations at a fixed value of lattice coverage $\theta=0,5$ given that, for symmetry arguments, they assumed that the critical coverage should be the same as the one for the monomer case, i.e. $\theta=0,5$. This assumption presupposes a symmetric phase diagram around $\theta=0,5$, a fact that as will be seen later, is not correct since the phase diagram is actually slightly asymmetric. For the determination of the critical temperature the authors employed a method based on the extrapolation of the maximum of the specific heat. Although this is a valid method, it is not as precise as others, such as the crossing of the fourth order Binder cumulants. In addition, taking into account that what is being looked for is actually a critical point ($T_c$ and $\theta_c$ and not only $T_c$), leaving the lattice coverage at the fixed value of 0.5 implies  that the critical temperature found will be incorrect if the true critical coverage is other than $\theta_c=0,5$. 

As was mentioned in section (3, simulations in the present work were run in the grand canonical ensemble with the aim of studying the problem in a wide range of both temperature and density with the chemical potential as a control parameter, and determine a precise critical point. The simulations for attractive dimers were run on square lattices of sizes $L=20,40,60,...,120$, with interaction energies $w=-1.40, -1.41, ..., -1.50$.

 At first, the dimer adsorption isotherms were obtained directly from the simulation for different lateral interaction energies as shown in Fig. 9. The figure highlights the typical jump that characterize a first order transition in a condensation-evaporation system. 
 
\begin{figure}

\includegraphics[scale=0.35]{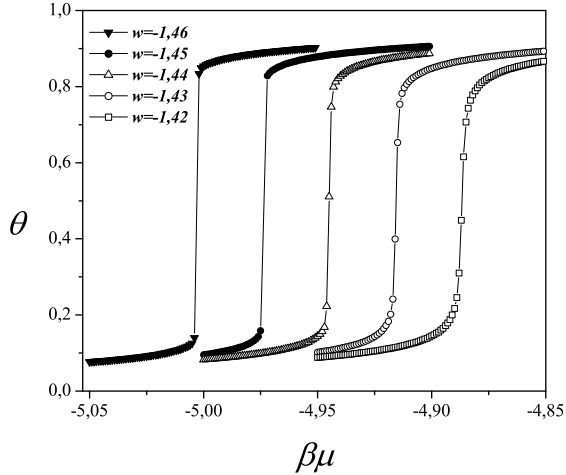}
\caption{Simulation adsorption isotherms for dimers. Typical behavior of the adsoprtion isotherms with different lateral interactions ($w$) near a first order phase transition. The isotherms corresponds to dimers ($k=2$) adsorbed on a square lattice for $w=-1,42$ (empty squares), $w=-1,43$ (empty circles), $w=-1,44$ (empty triangles), $w=-1,45$ (filled circles), $w=-1,46$ (filled triangles).}
\end{figure}

 Figure 10 shows the resulting Binder cumulant plot for the dimers case obtained by applying the technique presented in section (4). It can be seen from the figure a well located intersection point for the determination of the critical interaction energy $w_c=-1,444$ corresponding to a critical temperature $T_c=-1/w_c=0,692$. The critical temperature found here corrects the previous values obtained by other authors (Ramirez-Pastor et.al.  found $T_c=0,689$ \cite{SS3}) giving a  bit higher value than the preceding ones. 

\begin{figure}

\includegraphics[scale=0.35]{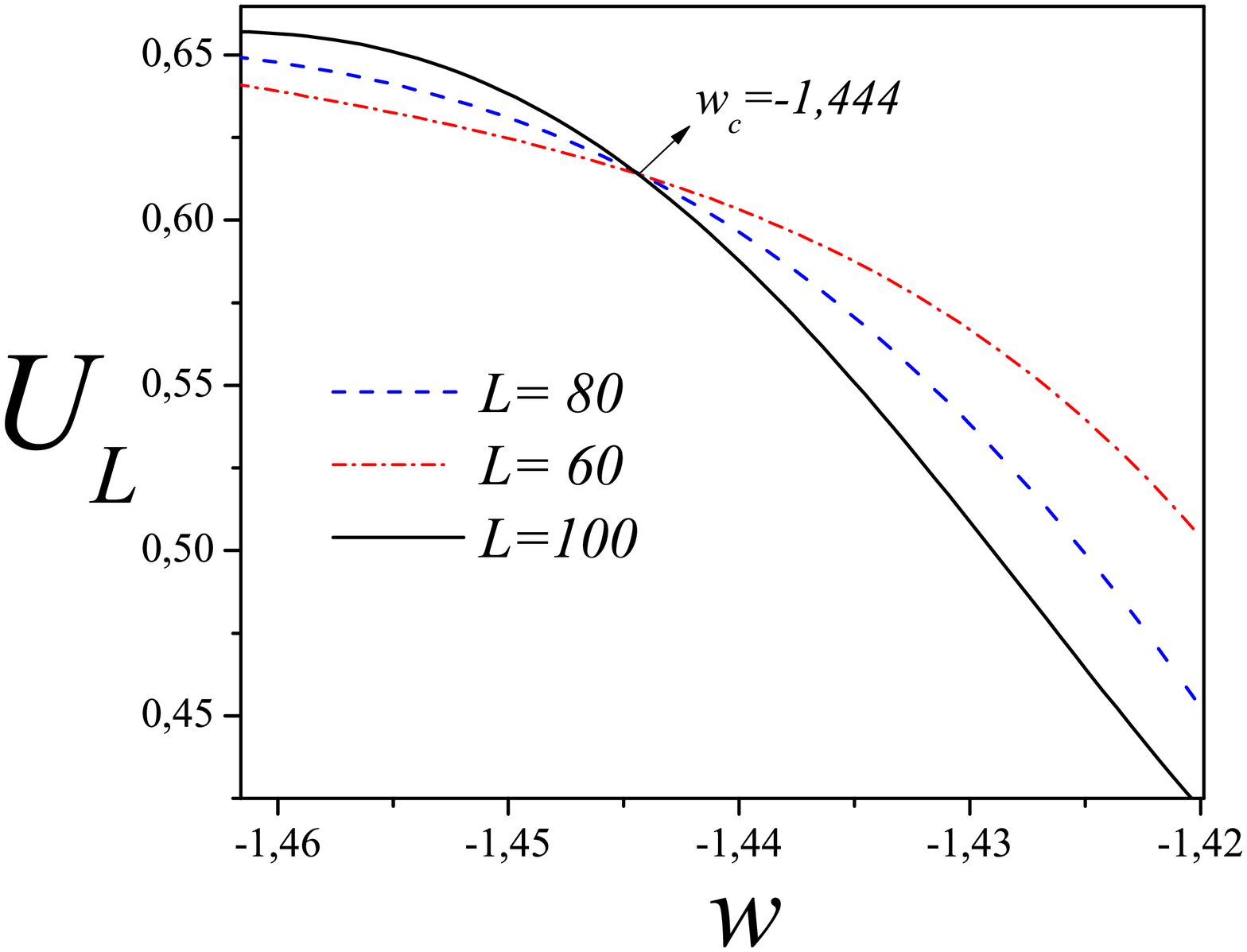}
\caption{Intersection plot of the fourth order Binder cumulant for the dimer adsorption system. The critical interaction energy found $w_c=-1,444$, is related to the critical temperature by $T_c=1/w_c=0.692$.}
\end{figure}

For the determination of the phase diagram for the dimer system we have employed, as was mentioned earlier in section (4), the two state approximation along with HR technique. The resulting phase diagram is shown in Fig. 11, where it can be seen that unlike the one obtained by Ramirez-Pastor et al. this phase diagram is slightly asymmetric. In addition, the critical coverage obtained in the present work is $\theta_c=0.489$ showing a small shift from the value $\theta_c^*=0.5$ predicted by previous works.

\begin{figure}

\includegraphics[scale=0.35]{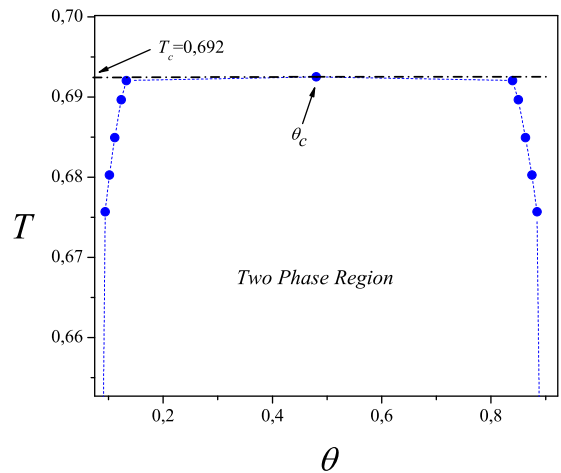}
\caption{Phase diagram for dimers. The dashed line indicates the critical temperature. It can be seen, the slightly asymmetric character of the diagram, resulting in a critical coverage different from 0.5 ($\theta_c =0.489$)}
\end{figure}

In summary, we can conclude that the method presented is adequate to find the critical temperature in a condensation-evaporation system, since it was successfully tested in a monomer system with attractive interactions, which has exact solution. The result obtained in this work was $T_c=0.56721$, while the exact value is $T_{c_{exact}}=0.56725$, exhibiting an error less than 1/1000. Using this method, the critical temperature for dimers with attractive interactions was found to be $T_c=0.692$, which was compared with Ramirez et al. \cite{SS3}, who reported a critical temperature $T_c=0.689$, which in light of the accuracy of the method  is clearly lower than the corresponding critical temperature reported in the present work. We believe that the value reported in this work is more accurate, not only because of the precision of the method used, but because Ramirez et al. \cite{SS3}, assume that the phase diagram is symmetric, so the critical density is assumed to be $\theta =0.5$. From our studies we have been able to conclude that the diagram is asymmetric and that the critical density is approximately $\theta_c =0.489$. The fact that the phase diagram is not symmetrical can be understood since there is no equivalence between occupied and empty sites in the case of dimers. The results obtained in the present work clearly correct and complement previous results in the literature.

\section{\label{sec:level1}ACKNOWLEDGMENTS}

 The authors thank support from Universidad Nacional de San Luis (Argentina) under project 322000, CONICET (Argentina) under project PIP 112-201101-00615 and the National Agency of Scientific and Technological Promotion (Argentina) under project PICT-2013-1678. The calculations were carried out using the BACO3 parallel cluster located at Instituto de F\'{\i}sica Aplicada, CONICET-Universidad Nacional de San Luis, Argentina.

\subsection{\label{sec:citeref}Citations and References}

\bibliographystyle{plain}

\bibliography{Manuscript}

\begin{thebibliography}{10}

\bibitem{Almarza}
No{\'e}~G Almarza, JM~Tavares, and MM~Telo da~Gama.
\newblock Effect of polydispersity on the ordering transition of adsorbed
  self-assembled rigid rods.
\newblock {\em Physical Review E}, 82(6):061117, 2010.

\bibitem{Binder}
Kurt Binder.
\newblock Applications of monte carlo methods to statistical physics.
\newblock {\em Reports on Progress in Physics}, 60(5):487, 1997.

\bibitem{RB40}
M~Borowko, W~R{\.z}ysko, and T~Staszewski.
\newblock Phase behavior of linear trimers confined to a narrow slit.
\newblock {\em Physical Review B}, 69(1):014209, 2004.

\bibitem{RB33}
Ma{\l}gorzata Bor{\'o}wko and Wojciech R{\.z}ysko.
\newblock Critical behavior of dimers in monolayers adsorbed on heterogeneous
  solid surfaces.
\newblock {\em Journal of colloid and interface science}, 244(1):1--8, 2001.

\bibitem{earl2005parallel}
David~J Earl and Michael~W Deem.
\newblock Parallel tempering: Theory, applications, and new perspectives.
\newblock {\em Physical Chemistry Chemical Physics}, 7(23):3910--3916, 2005.

\bibitem{ferrenberg1988new}
Alan~M Ferrenberg and Robert~H Swendsen.
\newblock New monte carlo technique for studying phase transitions.
\newblock {\em Physical review letters}, 61(23):2635, 1988.

\bibitem{ferrenberg1989optimized}
Alan~M Ferrenberg and Robert~H Swendsen.
\newblock Optimized monte carlo data analysis.
\newblock {\em Physical Review Letters}, 63(12):1195, 1989.

\bibitem{Fisher}
Michael~E Fisher.
\newblock Statistical mechanics of dimers on a plane lattice.
\newblock {\em Physical Review}, 124(6):1664, 1961.

\bibitem{Fowler}
RH~Fowler and GS~Rushbrooke.
\newblock An attempt to extend the statistical theory of perfect solutions.
\newblock {\em Transactions of the Faraday Society}, 33:1272--1294, 1937.

\bibitem{Hill}
Terrell~L Hill.
\newblock Statistical mechanics of adsorption. v. thermodynamics and heat of
  adsorption.
\newblock {\em The Journal of Chemical Physics}, 17(6):520--535, 1949.

\bibitem{Libro-Hill-2}
Terrell~L.. Hill.
\newblock {\em Statistical Mechanics: Principles and Selected Applications}.
\newblock Dover Publications, 1987.

\bibitem{HPT-3-hukushima}
Koji Hukushima and Koji Nemoto.
\newblock Exchange monte carlo method and application to spin glass
  simulations.
\newblock {\em Journal of the Physical Society of Japan}, 65(6):1604--1608,
  1996.

\bibitem{Kaste2}
Pieter~W Kasteleyn.
\newblock Dimer statistics and phase transitions.
\newblock {\em Journal of Mathematical Physics}, 4(2):287--293, 1963.

\bibitem{Kaste1}
PW~Kasteleyn.
\newblock The statistics of dimers on a lattice: I. the number of dimer
  arrangements on a quadratic lattice.
\newblock {\em Physica}, 27(12):1209--1225, 1961.

\bibitem{Lieb}
Elliott~H Lieb.
\newblock Solution of the dimer problem by the transfer matrix method.
\newblock {\em Journal of Mathematical Physics}, 8(12):2339--2341, 1967.

\bibitem{Lopez}
LG~L{\'o}pez, DH~Linares, and AJ~Ramirez-Pastor.
\newblock Comment on “effect of polydispersity on the ordering transition of
  adsorbed self-assembled rigid rods”.
\newblock {\em Physical Review E}, 85(5):053101, 2012.

\bibitem{muller1979monte}
H~M{\"u}ller-Krumbhaar and K~Binder.
\newblock Monte carlo methods in statistical physics.
\newblock {\em Topics in Current Physics}, 7, 1979.

\bibitem{Nagle}
John~F Nagle.
\newblock New series-expansion method for the dimer problem.
\newblock {\em Physical Review}, 152(1):190, 1966.

\bibitem{panagiotopoulos2000monte}
Athanassios~Z Panagiotopoulos.
\newblock Monte carlo methods for phase equilibria of fluids.
\newblock {\em Journal of Physics: Condensed Matter}, 12(3):R25, 2000.

\bibitem{JCP16}
PM~Pasinetti, F~Rom{\'a}, and AJ~Ramirez-Pastor.
\newblock First-order phase transitions in repulsive rigid k-mers on
  two-dimensional lattices.
\newblock {\em The Journal of chemical physics}, 136(6):064113, 2012.

\bibitem{Phares}
Alain~J Phares, Francis~J Wunderlich, David~W Grumbine, and Jonathan~D Curley.
\newblock The entropy curves for interacting dimers on a square lattice.
\newblock {\em Physics Letters A}, 173(4):365--368, 1993.

\bibitem{SS3}
AJ~Ramirez-Pastor, JL~Riccardo, and VD~Pereyra.
\newblock Monte carlo study of dimer adsorption at monolayer on square
  lattices.
\newblock {\em Surface science}, 411(3):294--302, 1998.

\bibitem{PRB4}
F~Rom{\'a}, AJ~Ramirez-Pastor, and JL~Riccardo.
\newblock Critical behavior of repulsive linear k-mers on square lattices at
  half coverage: Theory and monte carlo simulations.
\newblock {\em Physical Review B}, 68(20):205407, 2003.

\bibitem{PRB8}
F~Rom{\'a}, JL~Riccardo, and AJ~Ramirez-Pastor.
\newblock Critical behavior of repulsive dimers on square lattices at 2/ 3
  monolayer coverage.
\newblock {\em Physical Review B}, 77(19):195401, 2008.

\bibitem{RB38}
W~R{\.z}ysko and K~Binder.
\newblock Phase behaviour of heteronuclear dimers in three-dimensional
  systems—a monte carlo study.
\newblock {\em Journal of Physics: Condensed Matter}, 20(41):415101, 2008.

\bibitem{RB35}
W~R{\.z}ysko and M~Bor{\'o}wko.
\newblock Non-universal critical behaviour of heteronuclear dimers on a square
  lattice----a monte carlo study.
\newblock {\em Surface science}, 520(3):151--157, 2002.

\bibitem{RB34}
W~R{\.z}ysko and M~Bor{\'o}wko.
\newblock Phase diagrams of heteronuclear dimers adsorbed on a square lattice.
\newblock {\em The Journal of chemical physics}, 117(9):4526--4531, 2002.

\bibitem{RB39}
W~R{\.z}ysko and M~Bor{\'o}wko.
\newblock Computer simulation of phase diagrams of trimers adsorbed on a square
  lattice.
\newblock {\em Physical Review B}, 67(4):045403, 2003.

\bibitem{RB36}
W~R{\.z}ysko and M~Borowko.
\newblock Monte carlo study of adsorption of heteronuclear dimers on
  heterogeneous surfaces.
\newblock {\em Thin Solid Films}, 425(1):304--311, 2003.

\bibitem{RB37}
W~R{\.z}ysko and M~Bor{\'o}wko.
\newblock Phase behavior of unsymmetrical dimers on a square lattice.
\newblock {\em Surface science}, 600(4):890--896, 2006.

\bibitem{HPT-1-Swendsen}
Robert~H Swendsen and Jian-Sheng Wang.
\newblock Replica monte carlo simulation of spin-glasses.
\newblock {\em Physical Review Letters}, 57(21):2607, 1986.

\bibitem{HPT-4-2stateaprox}
Qiliang Yan and Juan~J de~Pablo.
\newblock Hyper-parallel tempering monte carlo: Application to the
  lennard-jones fluid and the restricted primitive model.
\newblock {\em The Journal of chemical physics}, 111(21):9509--9516, 1999.

\end{thebibliography}

\end{document}